\documentclass[namedreferences]{kluwer}

\usepackage{epsfig}

\def\mnras{MNRAS}
\def\aj{AJ}
\def\apj{ApJ}
\def\apss{ApSS}
\def\aaps{A\&ASS}
\def\pasp{PASP}

\begin{document}
\begin{article}

\begin{opening}



\title{Morphological classification of high redshift galaxies.}
\subtitle{GASPHOT: A tool for Galaxy Automated Surface PHOTometry.}

\author{E.
\surname{Pignatelli}\email{pignatel@sissa.it}\thanks{Partially 
supported by the Astronomical Observatory of Padova}}
\institute{SISSA/ISAS, Trieste, Italy}

\author{G. \surname{Fasano}\email{fasano@pd.astro.it}}
\institute{Osservatorio Astronomico di Padova, Padova, Italy}



\runningtitle{Morphology of high-z galaxies}
\runningauthor{E. Pignatelli and G. Fasano}

\begin{ao}
Ezio Pignatelli, c/o SISSA \\ 
e-mail: {\tt pignatel@sissa.it}; homepage: {\tt
http://www.sissa.it/\~\ pignatel/}
\end{ao}


\begin{abstract}
We present GASPHOT, a tool for automated surface photometry and
morphological classification of galaxies in deep and wide fields. The
requirements for any such tool are reviewed, and its use for the
classification of high-z galaxies is presented.
In the case if HDF-like images, for galaxies having a
magnitude ranging from  24 to 27.5, the uncertainties on the
photometric parameters derived from GASPHOT are respectively $\Delta M =
0.02-0.1$, $\Delta \log R_e \approx 0.03$, $\Delta n =0.02-0.5$.
 
\end{abstract}





\end{opening}


\vspace{-\baselineskip}
\section{Introduction}

In the last years the Hubble Space Telescope provided
very deep images of field galaxies, giving a glimpse of
their very first evolutionary phases. We now have a
chance to understand the basic elements of galaxy
formation and evolution, but we need new instruments to
handle the thousands of very faint objects that could be
present in one single image.

In particular, it has been noticed that at high redshift
- and for this kind of images - the Hubble classification
system could be, if not totally wrong, at least
not practical to use \cite{abraham96}. On the one hand,
the objects in the images are small, with a typical
radius of a few pixels. Thus, it is very hard to detect
the fine structure elements needed to distinguish the
different classes of spirals or to separate the barred
and unbarred families of galaxies (but see
\opencite{barred99}). On the other hand, the
morphological statistical analysis of the Hubble Deep
Fields \cite{abraham96} showed that at high redshift
(\hbox{$z\gsim 0.7$}) the percentage of Peculiar and
Irregular galaxies could be very high (up to 30\%).

Thus, the Hubble diagram does not appears very useful to
distinguish between the different classes of high
redshift galaxies. On the one hand, it brings plenty of
details for distinguishing between sub-classes of
galaxies (the spirals) that can not be separated for
distant objects; on the other hand, it gives few
informations about a class of galaxies (the Irregular)
which is dominant at high redshift, and that we would like
to split at least in truly irregular, peculiar
and interacting galaxies.

For these reasons, different authors tried to build new
classification systems, based on different {\em
quantitative} parameters. The most used is the luminosity
profile, measured either with a ``concentration
parameter'' \cite{abraham94} or with a slope of the
surface brightness profile; this can also be combined
with colors, asymmetry \cite{abraham96,conselice} or
power spectrum in Fourier space \cite{takamiya}.

In order to extract the valuable photometric parameters
(total magnitude, optical radius, $b/a,$ morphological
type $T$) for the thousands of objects expected in large
fields, we can not relieve on the usual photometric tool,
which are optimized to be used on single objects. Such
tools always involve some amount of interactivity, which
is ruled out by the large number of galaxies now
present. We need the photometric analysis process to be
completely automatic.
                 
While there are many tools available for the detailed
surface photometry of {\em single} galaxies, there are
few instruments which are build for the study of large
databases of objects (and they are {\bf not} for public
use). Usually the extraction and ``aperture photometry''
is handled by tools such as FOCAS \cite{focas} MORPHO
\cite{odewahn} or SExtractor \cite{sextractor}, while the
only software developed for automated surface photometry
of a large number of galaxies are GIM2D \cite{marleau}
and the HST MDS software \cite{ratnatunga}.

\vspace{-0.5cm}
\section{Structure and performances of GASPHOT}

For the reasons described above, we started creating a
photometric tool for detailed surface photometry of large
images.  The process is mainly divided in two steps:
first, a modified version of SExtractor \cite{sextractor}
is used to perform the identification and aperture
photometry of each object. Then a second program is used
to analyze the photometric profiles and derive the main
photometric parameters of each object.  Taking into
account the effects of the convolution with the PSF, the
aperture photometry is fitted with a Sersic law
$\mu \propto r^{1/n} $
with five free parameters: the total magnitude $M_{\rm tot}$,
the half-luminosity radius $R_e$, the Sersic index $n$
(which will be taken as a morphological index), the
flattening $b/a$ and the value of the local background.

The bias and errors in the determination of parameters
have been estimated by running IRAF simulations of
galaxies adopting the read-out noise, gain, background
level and PSF of the HST WFPC2 detector.

We performed simulations with galaxies having magnitude
from 23 to 27.5, and optical radii from 3 to 10
pixels. First, in order to remove the problems due to
blending effects, we analyzed the results of GASPHOT on a
sample of about 2000 galaxies positioned over a grid.
Galaxies were assumed to follow a pure exponential or de
Vaucouleurs law, but we also made a few tests with
galaxies following a Sersic law with $n$ ranging up to 6.
For galaxies having a magnitude of 24-27, we obtained
uncertainties of $\Delta M = 0.02-0.1$, $\Delta \log R_e
= 0.03$, $\Delta n = 0.02-0.5$, with no sizable bias.
Using the Hubble Deep Field image parameters, the limit
magnitude to obtain meaningful informations from the
photometry seems to be around 27.5.  The computational
time seems acceptable for images with thousands of
objects (about 20-30 seconds/object).

\begin{figure}[H]
\vspace{-0.5cm}
\centerline{\includegraphics[width=.35\textheight]{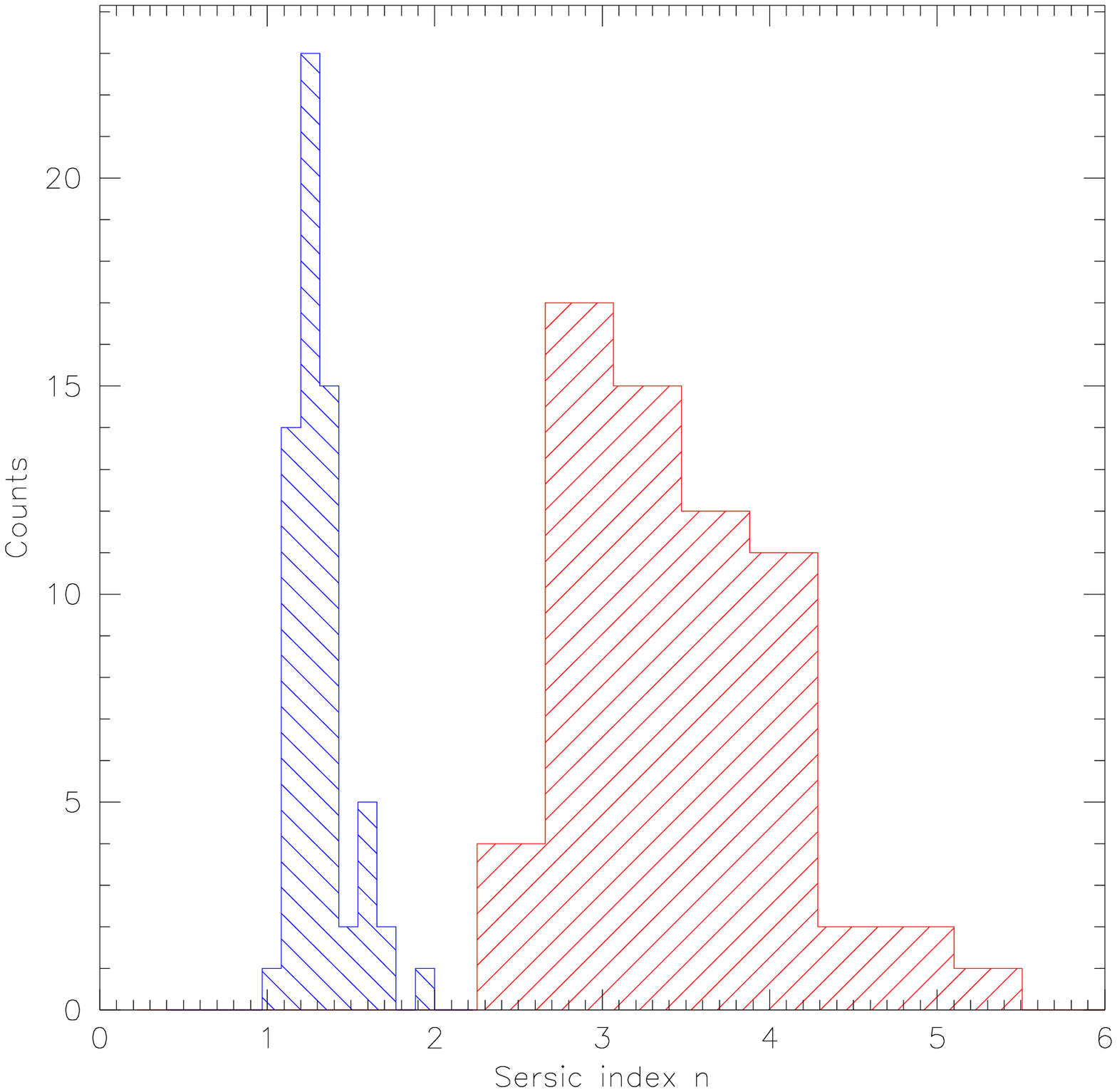}}
\caption{
Morphological classification of galaxies on grid, for
galaxies having total B magnitude of 27 in B. Note that
early- and late- type galaxies are clearly separated.
}
\label{fig:enne}
\end{figure}

The morphological classification of the objects has been
done on the basis of the Sersic index $n$ that best fits
the observations. In the future, we plan to include an
asymmetry parameter.  While the separations between
elliptical and spiral galaxies seems to be excellent
(see Fig.\ref{fig:enne}) we still have to test the
behaviour of the software when dealing with mixed types
such as S0s.

As a second, harder test, we spread 500 galaxies
uniformly with a random distribution over a 1400x1400
pixel image. The test included many galaxies heavily
blended or with a close companion. We also pushed our
magnitude limit up to 28.5, and tested our tool to
reproduce the original parameter of the galaxies. For
comparison, we tested the most common software,
SExtractor, on the same image. We only could

\begin{figure}[t]
\tabcapfont
\centerline{\centerline{
\begin{tabular}{cc}
\includegraphics[width=.35\textheight]{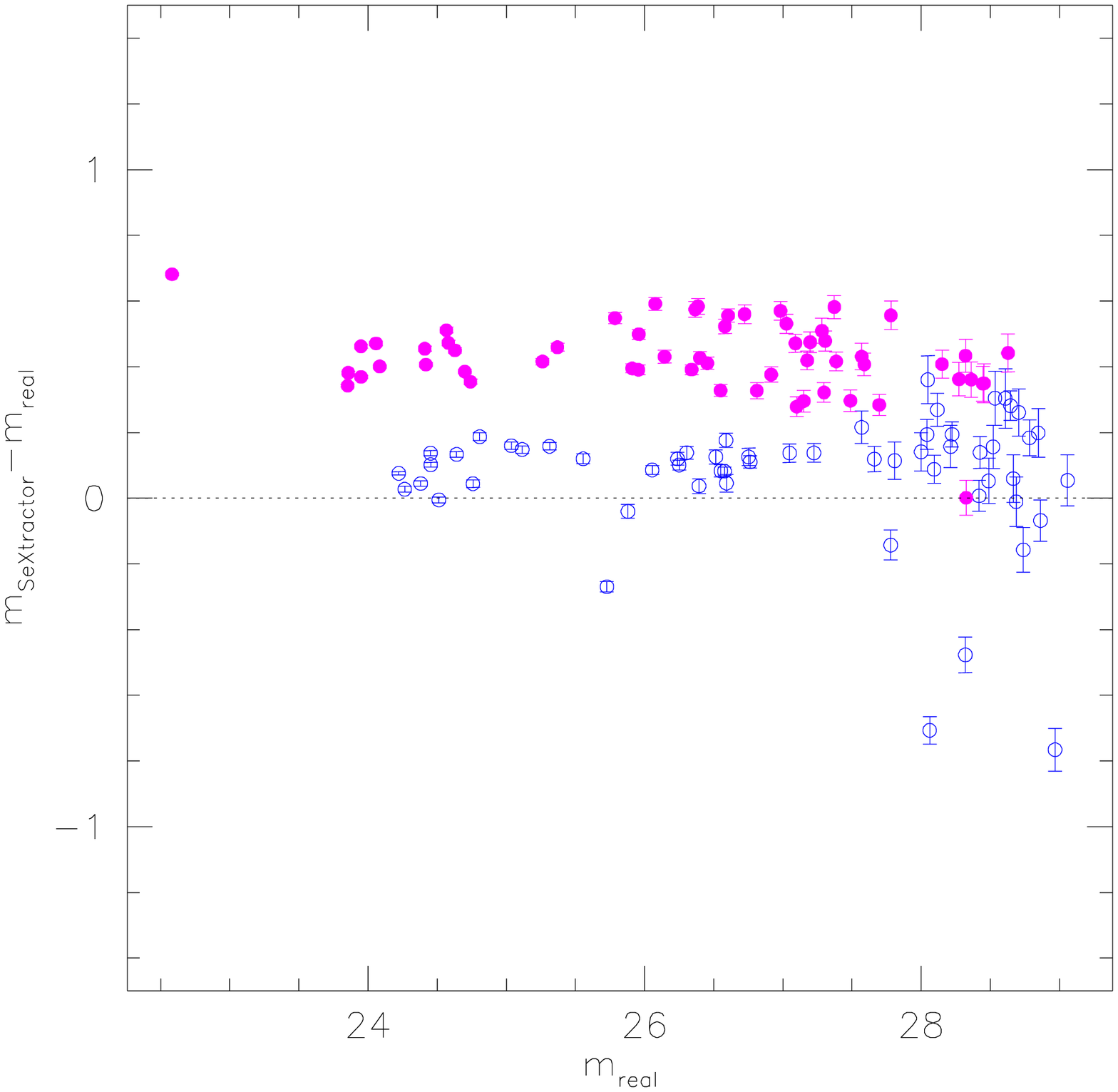} &
\includegraphics[width=.35\textheight]{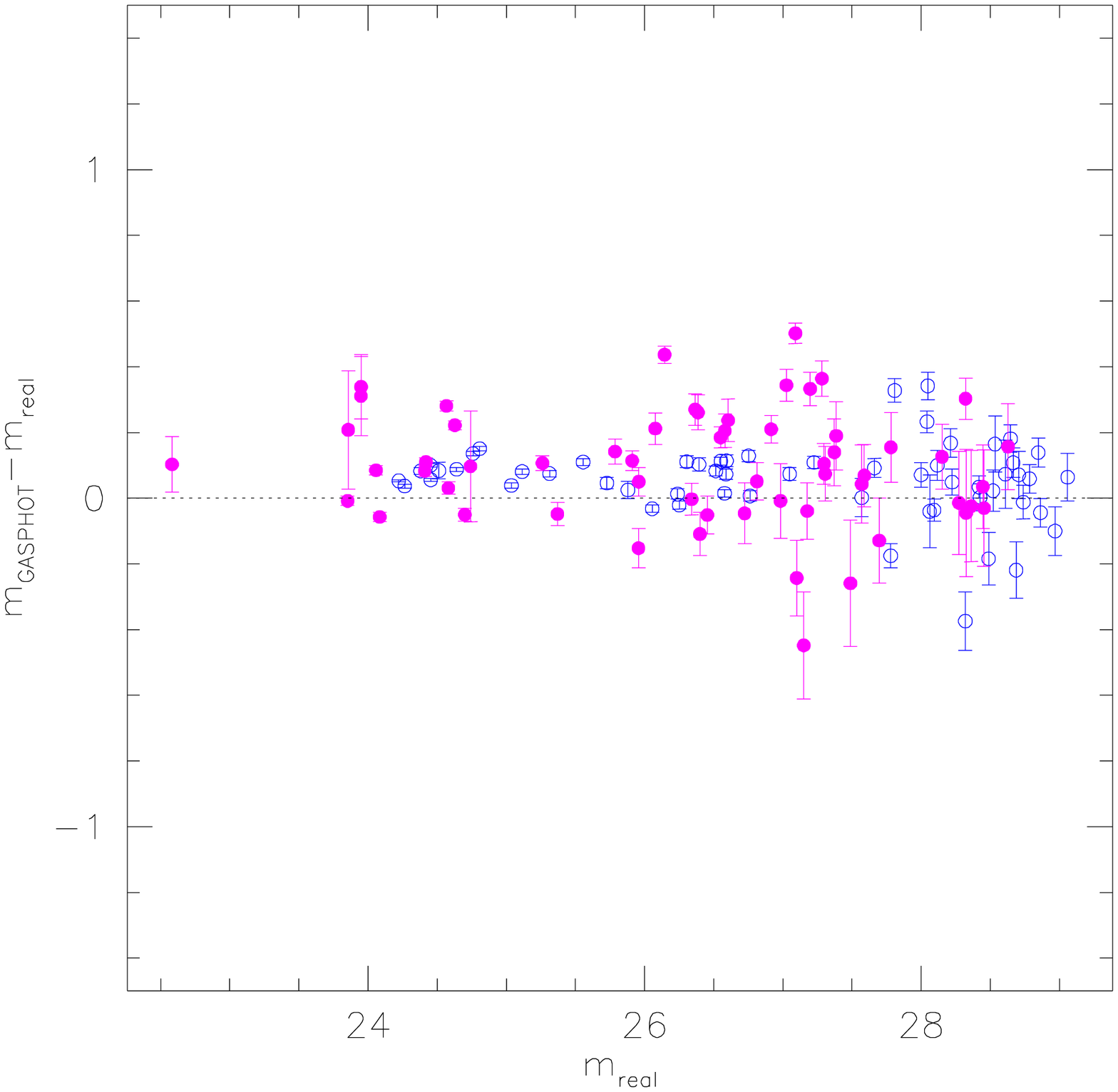} 
\end{tabular}}
\caption{Results for a simulated image of 500, uniformly
distributed, elliptical and spiral galaxies. On the left side we show the
disagreement of the magnitudes measured by SExtractor with the ``real''
magnitudes of the simulated galaxies. Full dots represent elliptical
galaxies, empty squares spiral galaxies. The same test is
performed on the right, for the same image, by GASPHOT.}}
\label{fig:mag}
\end{figure}
\vspace{-0.3cm}
\noindent
perform the test for magnitudes in this last case,
because SExtractor does not provide a morphological
classification or the radius $R_e$ (Fig.~\ref{fig:mag}).


In the future, we plan to produce a morphological and
photometric catalog of the Hubble Deep Fields using
GASPHOT, and to extend its application to a sample of
low-to-intermediate redshift clusters.  We also plan to
produce a public release of the software to be available
in the next few months in our web page.


{}
\vspace{-0.2cm}

\end{article}
\end{document}